\documentclass[11pt]{article}

\usepackage{amssymb, amsmath}
\usepackage{graphicx}
\usepackage{cite}

\newcommand{\const}{\,{\rm const}\,}

\newcommand{\cF}{\mathcal{F}}
\newcommand{\cA}{\mathcal{A}}
\def\be{\begin{equation}}
\def\ee{\end{equation}}
\def\bea{\begin{eqnarray}}
\def\eea{\end{eqnarray}}

\title{\bf{Uniqueness of near-horizon geometries of rotating extremal AdS$_4$ black holes}}

\author{Hari K. Kunduri$^a$\footnote{h.k.kunduri@damtp.cam.ac.uk } \  and James Lucietti$^b$\footnote{james.lucietti@durham.ac.uk } \\ \\
\small \sl $^a$DAMTP, University of Cambridge \\
\small \sl Centre for Mathematical Sciences, Wilbeforce Road, Cambridge, CB3 0WA, UK
\\ \\ \small \sl $^b$Centre for Particle Theory, Department of Mathematical Sciences, \\ \small \sl University of Durham, South Road, Durham, DH1 3LE, UK
 }

\date{}

\begin{document}

\maketitle

\begin{picture}(0,0)(0,0)
\put(350, 260){DCPT-08/67}
\put(350, 245){DAMTP-2008-114}
\end{picture}

\begin{abstract}

We consider stationary extremal black hole solutions of the
Einstein-Maxwell equations with a negative cosmological constant in
four dimensions.
We determine all non-static axisymmetric near-horizon geometries (with non-toroidal horizon topology) and all static near-horizon geometries for black holes of this kind. This allows us to deduce that the most general near-horizon geometry of an asymptotically globally AdS$_4$ rotating extremal black hole is the near-horizon limit of extremal Kerr-Newman-AdS$_4$. We also identify the
subset of near-horizon geometries which are supersymmetric.
 Finally, we show which physical quantities of extremal black holes may be computed from the near-horizon limit alone, and point out a simple formula for the entropy of the known supersymmetric AdS$_4$ black hole. Analogous results are presented in the case of vanishing
cosmological constant.
\end{abstract}

\section{Introduction}
Extremal black holes have been essential for uncovering the origin
of the Bekenstein-Hawking entropy from a statistical counting of
their microstates. It is well known that string theory first
reproduced the entropy formula, for a set of asymptotically flat
extremal black holes which are supersymmetric, by mapping the
gravitational system to a strongly-coupled 2d conformal field
theory~\cite{SV}. Restricting attention to supersymmetric states is
crucial as their degeneracy is not expected to change from weak to
strong coupling. The analogous calculation for asymptotically
AdS$_4$/AdS$_5$ supersymmetric black holes~\cite{KP,
Gutowski:2004ez, Gutowski:2004yv, Chong,Kunduri:2006ek}, where the
gauge-theory/gravity correspondence is used to map the problem to
one of enumerating appropriate operators in the dual
CFT$_3$/CFT$_4$, remains an important open problem. The AdS$_5$ case
is best understood and a certain amount of progress has been
made~\cite{KMMR, Berkooz1, Berkooz2, GGKM, Janik}. \par A
supersymmetric black hole is necessarily extremal, and there is
growing evidence to suggest that it is this latter property which is
behind the success of previous entropy calculations in flat space.
There are now a number of examples of successful microstate counting
for extremal, non-supersymmetric black holes~\cite{EH1, EM,
Horowitz:2007, Reall:2007jv, Emparan:2008qn}, including some recent
progress using asymptotic symmetries of near-horizon
geometries\cite{ Guica:2008mu, Hartman:2008pb, Azeyanagi:2008kb,
Lu:2008jk}. More generally, for rotating extremal black holes an
attractor mechanism has been demonstrated to exist under the
assumption that their associated near-horizon geometry must possess
an enhanced $SO(2,1)$ symmetry~\cite{Sen2,Sen:2007qy}. This
assumption was subsequently proved in a generic class of theories in
$D=4,5$ provided the black hole possesses $D-3$ rotational
symmetries~\cite{KLR2}\footnote{For analogous statements in vacuum
gravity in $D>5$ see~\cite{FKLR}.}. In particular, in~\cite{KLR2} it
was proved that such a near-horizon geometry must take the form
\begin{equation}
\label{canonical}
 ds^2 = \Gamma(\rho)\left[-C^2 r^2dv^2 + 2dvdr\right] + d\rho^2 + \gamma_{ij}(\rho)\left(dx^{i} + k^i r dv\right)\left(dx^{j} + k^j r dv\right)
\end{equation} where the horizon is at $r=0$, the Killing vector fields $\partial /\partial x^i$ generate the rotational symmetries ($i = 1, \dots, D-3$), $\Gamma$ and $\gamma_{ij}$ are smooth functions of the horizon coordinate $\rho$ and $C, k^i$ are constants.

The standard rigidity theorem in 4d, which states that a stationary, rotating, non-extremal black hole must be axisymmetric, has recently been generalised to extremal black holes~\cite{HI}. This implies the existence of one rotational symmetry and therefore {\it any} extremal rotating black hole in 4d must have a near-horizon limit given by (\ref{canonical}) . For black holes in $D\geq 5$ only one rotational symmetry has been proved to exist (for non-extremal~\cite{HIW} and for partial results in the extremal case~\cite{HI}) and thus the assumption of two in $D=5$ may constitute a genuine restriction on the space of solutions (we note that all known examples belong to this class)~\cite{R}.
\par It is remarkable that the classification problem for extremal black holes even in four dimensions remains unanswered. This is because the classic black hole uniqueness theorems assume that the horizon does not have degenerate components, i.e. the black hole is non-extremal. Moreover, these theorems are only valid for asymptotically flat spacetimes. Therefore, the classification of asymptotically AdS$_4$ black holes (extremal and non-extremal) also remains an open problem.

Extremality is a far weaker constraint than supersymmetry -- for
example, the vacuum Kerr black hole can be extremal but never
supersymmetric. While there are systematic techniques for
constructing supersymmetric solutions, it is also of interest to
develop techniques for classifying generic extremal black holes.
Indeed, every extremal black hole admits a near-horizon limit that
yields a geometry that solves the same theory. The field equations
reduce to a $D-2$ dimensional problem of Riemannian geometry on a
compact space. Classifying near-horizon geometries is thus a more
tractable problem and one can deduce important information about the
allowed extremal solutions in a given theory. The analysis reveals
not only the allowed horizon topologies but also their explicit
geometry. Hence, one may rule out the existence of black holes with
certain horizon topologies. The only drawback of using this
technique is that the existence of a near-horizon solution is not
sufficient to imply the existence of an extremal black hole with
that near-horizon geometry. However, the combination of a
near-horizon classification along with global information of the
black hole spacetime can provide a method to tackle the
uniqueness/classification problem for extremal black holes. Indeed,
such an approach was used to prove a uniqueness theorem for the
supersymmetric BMPV solution in five dimensional minimal ungauged
supergravity~\cite{R} and for the supersymmetric MP black holes of
four dimensional minimal ungauged supergravity~\cite{CRT2}.

It has turned out to be a difficult task to classify all near-horizon geometries in a given theory with no extra assumptions. For supersymmetric solutions this has been possible in 4/5/6d ungauged supergravity~\cite{CRT2,R,GMR}. For static near-horizon geometries it has also been possible in the vacuum (for any dimension and including the case of a cosmological constant)~\cite{CRT1} and 4d electrovacuum~\cite{CT}. However, in other cases it has been necessary to assume the existence of one or two rotational symmetries in 4d and 5d respectively. This assumption has allowed a classification of the near-horizon geometries of supersymmetric AdS$_5$ black holes in five-dimensional gauged supergravity~\cite{KLR1, KL1}. More recently, with the same assumptions, we performed such a classification for 4d (including a negative cosmological constant) and 5d  rotating extremal vacuum black holes~\cite{KL2}.

The purpose of this note is to perform a classification of all near-horizon geometries of stationary, extremal black hole solutions to four-dimensional Einstein-Maxwell theory including a non-positive cosmological constant. This theory is the bosonic sector of minimal (un)gauged supergravity, however we do not assume the solutions preserve any supersymmetry. Nevertheless, since these theories support supersymmetric black holes (which are necessarily extremal) our analysis will capture these as a subset to the space of all extremal solutions.
Our main results may be summarized as follows.
\paragraph{Result 1} Consider a four dimensional non-static, axisymmetric near-horizon geometry with a compact horizon section of non-toroidal topology, satisfying the Einstein-Maxwell equations with cosmological constant $\Lambda$. If $\Lambda =0 $ then it must be the near-horizon geometry of the rotating extremal Kerr-Newman black hole. If $\Lambda < 0$ it must be the near-horizon geometry of the rotating extremal Kerr-Newman-AdS$_4$.

\paragraph{Result 2} Consider a four dimensional static near-horizon geometry with a compact horizon section, satisfying the Einstein-Maxwell equations with cosmological constant $\Lambda$. For $\Lambda <0$ it must be a direct product of $AdS_2$ with a  metric of constant curvature on $S^2$, $T^2$, or (compact quotients of) $H^2$. For $\Lambda = 0$ it must be a direct product of $AdS_2$ and $S^2$. \\
\par \noindent

\noindent {\bf Remarks}
\begin{itemize}
\item These results do not employ any asymptotic information of the relevant black hole solutions. For $\Lambda=0$ we are of course interested in asymptotically flat black holes. For $\Lambda<0$ we are mostly interested in asymptotically {\it globally AdS }solutions. In these cases topological censorship~\cite{top} immediately implies that the horizon topology is $S^2$.
\item In~\cite{HI} it has been shown that an asymptotically flat or globally AdS extremal rotating black hole must be axisymmetric\footnote{Although this theorem is only stated explicitly for the asymptotically flat case, it also applies to asymptotically globally AdS black holes. In fact, the only parts of the proof which employ the asymptotic information, are where it is used to establish that the stationary Killing field does not vanish on the event horizon (so one can define a spacelike foliation of the event horizon) and that the exterior is simply connected (topological censorship). Both of these properties are still true for asymptotically globally AdS black holes.}. It follows that its near-horizon limit, which may be non-static or static, must be axisymmetric. If non-static our Result 1 implies that it must be given by the near-horizon geometry of the $J \neq 0$ extremal Kerr-Newman-(AdS$_4)$. If static Result 2 and topological censorship (which tells us the horizon must be $S^2$) implies it is given by $AdS_2 \times S^2$ which is the near-horizon geometry of extremal Reissner-Nordstrom-(AdS$_4$) (which is the $J=0$ extremal Kerr-Newman-(AdS$_4$)).
\item It has been shown that asymptotically flat, non-extremal, non-rotating black holes must be static~\cite{SW, W}. If this result extends to extremal and AdS black holes then our Result 2 shows it must be given by the near-horizon limit of extremal Reissner Nordstrom-(AdS$_4$).
\item The $\Lambda=0$ case of Result 1 has been proved in the context of extremal isolated horizons~\cite{LP}. The classification of extremal isolated horizons is in fact mathematically equivalent to that of near-horizon geometries. Also see~\cite{liko} for some results on the AdS case.
\item The near-horizon geometry of the supersymmetric AdS$_4$ black hole~\cite{KP, CK0} is non-static and hence given by a subset of the solutions in Result 1. In fact we will show this subset is the most general non-static, axisymmetric, supersymmetric near-horizon geometry (with non-toroidal horizon topology). We will also present a simple formula for their entropy.
\item Many of our proofs are also valid for $\Lambda>0$. Indeed our method determines the near-horizon geometry in this case too, although we have not analysed the resulting solutions in detail.
\end{itemize}
This note is organized as follows. First we set up the near-horizon
equations which must be satisfied by a near-horizon geometry
solution to the Einstein-Maxwell equations with a cosmological
constant. Next, we determine all static near-horizon geometries. We
then determine all non-static axisymmetric near-horizon geometries.
We then identify which subset of the derived near-horizon geometries
are supersymmetric. Finally, we present general formulas for the
electric and magnetic charges and the angular momentum of extremal
black hole solutions to this theory, written as integrals of the
near-horizon data over the horizon.

\section{Electro-vacuum near-horizon equations}
We consider solutions of $D=4$ Einstein-Maxwell theory with a cosmological constant. The field equations are
\begin{eqnarray}\label{EOM}
R_{\mu\nu} &=& T_{\mu\nu} \equiv 2\left(\cF_{\mu}^{\phantom{a}\delta}\cF_{\nu \delta} - \frac{1}{4}g_{\mu\nu}\cF^2\right) + \Lambda g_{\mu\nu} \\
d\star \mathcal{F} &=& 0, \qquad d\cF =0 \nonumber
\end{eqnarray}
where $\cF$ is the Maxwell two form and we write $\cF=d\cA$.
We will be mainly interested in the cases $\Lambda=0$ and $\Lambda<0$ which correspond to the bosonic sectors of minimal ungauged or gauged supergravity (with gauge coupling given by $\Lambda=-3g^2$) respectively, although many of our proofs also work for $\Lambda>0$.

The event horizon of a four dimensional stationary extremal black hole (asymptotically flat or AdS) must be a Killing horizon of a Killing vector field $V$~\cite{HI}. In a neighbourhood of such a Killing horizon we can always introduce Gaussian  null coordinates $(v,r,x^a)$ such that $V = \partial / \partial v$, the horizon is at $r=0$ and $x^a$ ($a=1,2$) are coordinates on $\mathcal{H}$, a spatial section of the horizon. Note that $\mathcal{H}$ is a two-dimensional compact manifold without boundary. The black hole metric and Maxwell field in these coordinates are
\bea
 ds^2 &=& r^2 F(r,x) dv^2 + 2dvdr + 2rh_a(r,x) dvdx^a + \gamma_{ab}(r,x)dx^a dx^b \\
\label{MaxGau}
\cF &=& \cF_{vr}dv\wedge dr + \cF_{ra}dr \wedge dx^{a}  + {\cF}_{va} dv \wedge
dx^a  +\frac{1}{2}\cF_{ab}dx^{a} \wedge dx^{b} \; .
\eea
The near-horizon limit \cite{R,KLR2} is obtained by taking the limit $v \to v/\epsilon, \ r \to \epsilon r$ and $\epsilon \to 0$. The resulting metric is
\begin{equation}\label{gNH}
 ds^2 = r^2F(x)dv^2 + 2dvdr + 2rh_a(x)dv dx^a + \gamma_{ab}(x)dx^a dx^b
\end{equation} where $F, \ h_a, \ \gamma_{ab}$ are a function, a one-form, and a Riemannian metric respectively, defined on $\mathcal{H}$. In general the Maxwell field~(\ref{MaxGau}) does \emph{not} admit a near-horizon limit due to the $\cF_{va}$ component. However, we can use the field equations to show that it must for solutions. It is well known that for a  Killing horizon $\mathcal{N}$ of $\xi$ one must have $R_{\mu\nu}\xi^{\mu}\xi^{\nu}|_{\mathcal{N}}=0$. Taking $\mathcal{N}$ to be the event horizon with $\xi=V$, and using~(\ref{EOM}) one finds:
\be
R_{\mu\nu}\xi^{\mu}\xi^{\nu}|_\mathcal{N}= 2\gamma^{ab}
\cF_{va}\cF_{vb}|_{r=0}
\ee which implies $\cF_{va}=0$ at $r=0$. It follows (assuming analycity) that
$\cF_{va}=r \hat{\cF}_{va}$ for some regular functions
$\hat{\cF}_{va}$. This guarantees that the near-horizon limit of the Maxwell field always exists, and is given by:
\be
\cF=\cF_{vr}(x)dv \wedge dr + r\hat{\cF}_{va}(x) dv \wedge dx^a +
\frac{1}{2}\cF_{ab}(x)dx^a \wedge dx^b  \; .\ee
Note that the Bianchi identity $d\cF=0$ further constrains the Maxwell field and implies it can be written as
\be\label{FNH}
\cF=d(\Delta(x) rdv)+ \hat{\cF}
\ee
where $\hat{\cF}\equiv \frac{1}{2}\cF_{ab}(x)dx^a \wedge dx^b$ is a closed two-form and $\Delta \equiv -\cF_{vr}$ is a function, both defined on $\mathcal{H}$. Note that we can locally introduce a potential $\hat{A}$ on $\mathcal{H}$ such that $\hat{\cF}=d\hat{A}$.
\par The purpose of this note is to determine all electrovacuum extremal black hole near-horizon geometries with a cosmological constant in four dimensions. This is equivalent to finding the most general metric and Maxwell field of the form~(\ref{gNH}) and (\ref{FNH}) that satisfy~(\ref{EOM}). A lengthy calculation reveals that the spacetime field equations are equivalent to the following set of equations on $\mathcal{H}$:
\begin{eqnarray}
 R_{ab} &=& \frac{1}{2}h_a h_b - \nabla_{(a}h_{b)} + \Lambda \gamma_{ab} + 2\hat{\cF}_{ac}\hat{\cF}_{bd}\gamma^{cd} + \Delta^2\gamma_{ab} - \frac{\gamma_{ab}}{2}\hat{\cF}^2 \label{Riceq} \\
F &=& \frac{1}{2}h_ah^a - \frac{1}{2}\nabla_a h^a + \Lambda -\Delta^2 - \frac{\hat{\cF}^2}{2} \label{Feq} \\
\label{4dmaxwell}
d\star_2 \hat{\cF} &=& \star_2 i_h\hat{\cF}+ \star_2 (d\Delta-\Delta h).
\eea 
where $R_{ab}$, $\nabla$ and $\star_2$ are the Ricci tensor, the covariant derivative and Hodge dual of the 2d metric $\gamma_{ab}$. In particular, (\ref{Riceq}) is the $ab$ component of the Einstein equations, (\ref{Feq}) is the $vr$ component and (\ref{4dmaxwell}) is the Maxwell equation, all written covariantly on $\mathcal{H}$. It can be shown that the rest of the Einstein equations are satisfied as a consequence of the above set of equations. It is useful to note that using equations (\ref{Riceq}), (\ref{Feq}) and (\ref{4dmaxwell}), it can be shown that the contracted Bianchi identity $\nabla^b(R_{ab}- \frac{1}{2}\gamma_{ab}R)=0$  is equivalent to
\begin{equation}
\nabla_a F =  Fh_a +2h_b{\nabla}_{[a}h_{b]} - {\nabla}_{b}{\nabla}_{[a}h_{b]}  -2(\hat{\cF}_{ab}+\Delta \gamma_{ab})(\nabla^b \Delta -\Delta h^b) \label{gradF} \; .
\end{equation}

\section{Static near-horizon geometries}
A static near-horizon geometry is one for which the Killing field
normal to the horizon is hypersurface orthogonal, i.e. $V \wedge dV
=0$ everywhere. Such solutions have been classified previously in
the vacuum (including a negative cosmological constant)~\cite{CRT1}
and electrovacuum (with no cosmological constant)~\cite{CT} and
considered more generally in~\cite{KLR2}. In~\cite{KLR2} it was
shown that staticity is equivalent to the following constraints on
the metric: $dF = Fh$ and $dh=0$. We first derive an analogous
constraint for the Maxwell field. Defining the twist one-form
$\omega = \frac{1}{2} \star(V \wedge dV)$, one can check $d\omega =
-\star(V \wedge R(V))$ where $R(V)_{\mu} = R_{\mu\nu}V^{\nu}$.
Therefore a static near-horizon must be Ricci-static, i.e. $V \wedge
R(V)=0$. From~(\ref{EOM}) it follows that
\begin{equation}
 V \wedge R(V) = 2V \wedge \cF_{\mu\rho}\cF^{\rho\nu}V_{\nu} dx^{\mu}.
\end{equation} From this it is easy to check that a near-horizon geometry is Ricci static if and only if $d\Delta =  h\Delta$. It follows that a static near-horizon geometry must have $d\Delta =  h\Delta$.

We now turn to solving the staticity conditions. Since $dh=0$, we see that {\it locally} we can always write $h=d\lambda$. If $\mathcal{H}$ is simply connected (as for $S^2$) then the function $\lambda$ is actually globally defined. If $\mathcal{H}$ is not simply connected, consider an open cover of simply connected sets $\{ U_i \}$; then in each set $U_i$ one can write $h=d\lambda_i$. Defining $\psi=e^{-\lambda /2}$, we can integrate (in each $U_i$) the remaining staticity conditions to get $F = F_{0}^i\psi_i^{-2} $ and $\Delta=e_i \psi_i^{-2}$ for some constants $F_{0}^i$ and $e_i$.

Now consider the near-horizon Maxwell equation~(\ref{4dmaxwell}). This simplifies considerably to
\be
d\phi-\phi d\lambda=0
\ee
where $\phi= \star_2 \hat{\cF}$ is a (globally defined) function on $\mathcal{H}$. It follows that (in each $U_i$) $\phi= b_i\psi_i^{-2}$.

We may now deduce an important fact.  First note that if $e_i=b_i=0$ for all $i$ then $\cF=0$ everywhere on $\mathcal{H}$ which is the vacuum case studied in~\cite{CRT1}. Therefore, for a non-trivial Maxwell field we must have at least one of $e_i$ or $b_i$ non-zero, which we will assume. By comparing the expressions for $\Delta$ and $\phi$ on the overlaps $U_i \cap U_j$  we must have $e_i \psi_i^{-2}=e_j\psi_j^{-2}$ and  $b_i\psi_i^{-2} =  b_j\psi_j^{-2}$. We deduce that either all the $e_i$ are non-zero (and have the same sign) or all the $b_i$ are non-zero (and have the same sign). Using the fact that the $\psi_i$ are only defined up to a multiplicative constant depending on $i$ (since $\lambda_i$ are defined up to an additive constant depending on $i$) we may set $e_i=e_j$ (if these are non-zero) or $b_i=b_j$ (if $e_i=0$) to deduce $\psi_i=\psi_j$. It follows that $F_0^i=F_0^j$ and $b_i=b_j$. Thus we deduce that $\psi$ is a globally defined function even in the non simply connected case and from now on we drop all indices $i$ and work on the whole of $\mathcal{H}$. Note we have determined: $F=F_0\psi^{-2}$, $\Delta=e\psi^{-2}$ and $\phi= b \psi^{-2}$.

We now turn to solving the rest of the equations on $\mathcal{H}$.  Our analysis closely parallels that of~\cite{CRT1}. First note that~(\ref{gradF}) is automatically satisfied as a consequence of staticity. Then~(\ref{Riceq}) and~(\ref{Feq}) can be written as
\bea
&& \psi R_{ab}= 2\nabla_a\nabla_b \psi + \psi T_{ab}, \label{Ricstat}\\
&& F_0 = \psi\nabla^2\psi + | \nabla \psi |^2 +\psi^2 T_{vr}
=\frac{1}{2}\nabla^2\psi^2 +\psi^2 T_{vr} \label{Fstat} \eea where
\be
T_{ab}= (e^2+b^2)\psi^{-4} \gamma_{ab} +\Lambda\gamma_{ab}, \qquad T_{vr}= -(e^2+b^2)\psi^{-4}+\Lambda \; .
\ee
Following the procedure given in~\cite{CRT1}, which uses the fact
that $\mathcal{H}$ is two-dimensional so $R_{ab} =
\frac{1}{2}g_{ab}R$, allows one to show that \be
\label{Ricsca1}\psi^3R= 6(e^2+b^2)\psi^{-1}+\frac{2\Lambda
\psi^3}{3}+c \ee where $c$ is a constant. Taking the trace
of~(\ref{Ricstat}) gives $2\nabla^2\psi =\psi(R-\gamma^{ab}T_{ab})$
and substituting~(\ref{Ricsca1}) gives \be \label{lappsi}
2\nabla^2\psi= c{\psi}^{-2} + 4(e^2+b^2)\psi^{-3} -\frac{4\Lambda
\psi}{3}\ee There are two cases to consider: either (i) $d\psi \neq
0$ in some open set in $\mathcal{H}$, or (ii) $\psi$ is constant in
$\mathcal{H}$.

 Let us treat case (i) first. Since $d\psi \neq 0$ in some open set, we can use $\psi$ as a coordinate there and introduce another
coordinate $\varphi$ such that \be \gamma_{ab}dx^a dx^b =
\frac{d\psi^2}{| \nabla \psi|^2} +H(\psi,\varphi) d\varphi^2 \; .
\ee To calculate $|\nabla \psi |^2$ we proceed as in~\cite{CRT1} and
take the divergence of~(\ref{Ricstat}) leading to \be | \nabla \psi
|^2 +\frac{\psi^2 R}{2}-2(e^2+b^2)\psi^{-2} = \alpha \ee for some
constant $\alpha$. Using our expression for the Ricci scalar then
gives \be \label{Pdef} | \nabla \psi |^2 = \alpha -\frac{c}{2\psi}-
\frac{e^2+b^2}{\psi^2}-\frac{\Lambda \psi^2}{3} \equiv P(\psi) \; .
\ee The Laplacian in $(\psi,\varphi)$ coordinates is \be \nabla^2
\psi =P'(\psi) + P\frac{\partial }{\partial \psi}\log
\sqrt{\frac{H}{P}} \ee  which implies that equation (\ref{lappsi})
becomes \be \frac{\partial}{\partial \psi} \log \sqrt{\frac{H}{P}}
=0 \ee and thus $H(\psi,\phi)=g(\phi)P(\psi)$ for some function
$g(\phi)$. Therefore, introducing a coordinate
$d\tilde{\varphi}^2=gd\varphi^2$ the metric on $\mathcal{H}$ is
simply \be \label{nonconstpsi}\gamma_{ab}dx^adx^b=
\frac{d\psi^2}{P(\psi)}+P(\psi) d\tilde{\varphi}^2 \ee where
$P(\psi)$ is given by (\ref{Pdef}). If one changes coordinates $r
\to \psi^2 r$, then the full near-horizon geometry simplifies to
\bea
ds^2 &=& \psi^2( F_0 r^2 dv^2 +2dvdr)+  \frac{d\psi^2}{P(\psi)}+P(\psi) d\tilde{\varphi}^2 \\
\cF &=& e dr \wedge dv + b \psi^{-2} d\psi \wedge d\varphi \eea It
is worth pointing out that this near-horizon solution can be
obtained by an analytic continuation of the full non-extremal
Reissner-Nordstrom-AdS black hole solution, and can be non-singular
for non-compact $\mathcal{H}=R^2$. However, for compact
$\mathcal{H}$ we can show that the horizon metric must be singular.
To see this, write $P=\psi^{-2}Q$ where $Q$ is a quartic in $\psi$.
Compactness requires $d\psi$ to vanish at distinct maxima and minima
of $\psi$ and thus, noting $(d\psi)^2=\psi^{-2}Q(\psi)$, we see that
$Q(\psi)$ must have real roots at these points $\psi_1<\psi_2$ with
$\psi_1 \leq \psi \leq \psi_2$ and $Q(\psi)>0$ inside the interval.
The condition for the absence of conical singularities at these
endpoints implies $Q(\psi)$ must be even (see~\cite{KL2} for an
identical argument), i.e. $c=0$ and thus $\psi_1=-\psi_2<0$. However
by definition $\psi>0$ and therefore we have contradiction and hence
one cannot simultaneously remove the conical singularities. We
therefore rule out this case.

Now consider case (ii), i.e. the case when $\psi$ is a constant. It
follows that $\lambda$ is constant, so $h=0$.
Then we see that $F$, $\Delta$ and $\phi$ are constants and
$F=-\Delta^2-\phi^2 +\Lambda$. Observe that the full near-horizon
geometry is then\bea \label{staticNH}
ds^2 &=& (-\Delta^2-\phi^2+\Lambda)r^2dv^2 +2dvdr + \gamma_{ab}dx^a dx^b\\
\cF &=& \Delta dr \wedge dv + \phi \star_2 1 \nonumber\eea and since
$F=-\Delta^2-\phi^2+\Lambda<0$ (for $\Lambda \leq 0$ and $\cF \neq
0$) it is simply the direct product $AdS_2 \times \mathcal{H}$. The
remaining equation is \be R_{ab}= (\Delta^2+\phi^2+\Lambda)
\gamma_{ab} \ee which implies the metric $\gamma_{ab}$ is locally
isometric to one of the maximally symmetric metrics on $S^2,T^2,H^2$
depending on the sign of $\Delta^2+\phi^2+\Lambda$.

Topological censorship implies that in the case of asymptotically
flat or globally AdS space-times,
then $\mathcal{H}=S^2$. We have shown that the only {\it regular} static near-horizon geometry in
this case is a direct product of $AdS_2 \times S^2$, given by
(\ref{staticNH}), parameterised by constants $\Delta, \phi$ which
satisfy $\Delta^2+\phi^2+\Lambda>0$.  This solution is simply the near-horizon limit of extremal
Reissner-Nordstrom-AdS$_4$. Furthermore, in the asymptotically locally AdS case, where higher genus horizons are allowed, we have shown that the only static near-horizon geometries are given again by (\ref{staticNH}) with $\Delta^2+\phi^2+\Lambda=0$ for $\mathcal{H}=T^2$, or $\Delta^2+\phi^2+\Lambda<0$ for $\mathcal{H}=\Sigma_g$.

\section{Non-static near-horizon geometries}
It has been shown that a 4d stationary rotating extremal black hole must be axisymmetric (for both asymptotically flat and globally AdS)~\cite{HI}. Thus, we assume the existence of a rotational Killing vector field $m$ for the full black hole spacetime, with closed spacelike orbits. The near-horizon limit of such a black hole inherits this symmetry and therefore we need only solve for the most general axisymmetric near-horizon geometry.  It follows that $m$ leaves the near-horizon data $(F,h_a,\gamma_{ab}, \Delta, \hat{\cF})$ invariant. Therefore $m$ is also a Killing vector field of the horizon metric. It should be noted that a compact 2d manifold with a global $U(1)$ action must be topologically $S^2$ or $T^2$ and therefore $\mathcal{H}$ can only have these topologies~\cite{Gowdy}. As in the vacuum case\cite{KL2}, some of our global arguments do not apply to toroidal horizons and therefore we will assume $S^2$ horizon topology.\footnote{It should be noted that for $\Lambda=0$ it is easy to show that the only $T^2$ case (even without assuming axisymmetry) is the flat solution $R^{1,1} \times T^2$ with vanishing Maxwell field. To see this note that using (\ref{Riceq}), the vanishing of the Euler number gives $0=\int_{\mathcal{H}} R= \int_{\mathcal{H}} \left(\frac{1}{2} h^2 + \hat{\cF}^2+2\Delta^2 \right)$, which implies $h \equiv0$, $\hat{\cF} \equiv 0$ and $\Delta \equiv 0$. This in turn implies that $F \equiv 0$ and the metric $\gamma_{ab}$ is flat. This argument fails when $\Lambda<0$ and it is an interesting problem to classify solutions in this case (even assuming axisymmetry).} Note that for $S^2$ topology, $m$ must vanish at two distinct points (the poles).

We introduce coordinates $(\rho,x)$ on $\mathcal{H}$ adapted to this symmetry, so $m=\partial /\partial x$, and
\bea
\label{2du1}\gamma_{ab}dx^adx^b &=& d\rho^2+\gamma(\rho) dx^2, \qquad h=\Gamma^{-1}[ \gamma(\rho) k(\rho)dx-\Gamma'd\rho ] \\
\hat{\cF} &=& B(\rho)d\rho \wedge dx
\eea
which define the functions $(\gamma,k,\Gamma,B)$ with $\Gamma>0$ and $f'(\rho)=df/d\rho$.  These coordinates are then valid everywhere on $\mathcal{H}$ except at the points where the rotational Killing field $m$ vanishes, where a separate analysis of regularity will be performed.

It is useful to first consider the near-horizon Maxwell equation (\ref{4dmaxwell}). It is equivalent to two equations, corresponding to the $x$ and $\rho$ components of (\ref{4dmaxwell}), which are:
\be
\label{u1max}
kB= (\Delta \Gamma)', \qquad \left(\frac{B\Gamma}{\sqrt{\gamma}}\right)'+k\Delta \sqrt{\gamma}=0 \; .
\ee
respectively.

For completeness we will now show how the symmetry enhancement result we proved in~\cite{KLR2} emerges in the present formalism. Firstly, note that $T_{\rho x}=0$, so equation (\ref{Riceq}) gives $R_{\rho x}= -\frac{1}{2}\Gamma^{-1}\gamma k'$ and since $R_{\rho x}=0$ automatically for a metric of the form (\ref{2du1}), we see that the function $k$ must be constant. Now, the $\rho$ component of (\ref{gradF}) is
\be
F'=-\frac{\Gamma'}{\Gamma}F+\Gamma^{-1}k(\Gamma^{-1}\gamma k)' +2\Delta \Gamma^{-1}(kB-(\Delta\Gamma)')
\ee
which using the $x$ component of the Maxwell equation is identical to the vacuum case. Defining $A=\Gamma^2F-\gamma k^2$ then implies $A'+\frac{\Gamma'}{\Gamma}A+ k'k=0$ and thus using the fact that $k'=0$ we learn that $A=A_0\Gamma$ for constant $A_0$. This is sufficient to prove the near-horizon geometry symmetry enhancement result of~\cite{KLR2}. This can be seen by changing coordinates $r \to \Gamma(\rho)r$, in which case the metric and Maxwell field simplify to
\bea
ds^2 &=& \Gamma(\rho)[A_0 r^2dv^2+2dvdr]+ d\rho^2+\gamma(\rho)(dx+krdv)^2 \\
\cF &=& E dr \wedge dv +B d\rho \wedge (dx+rdv)
\eea
where we have defined $E \equiv \Delta \Gamma$ and used $kB=E'$ to rewrite the Maxwell field.
Note that $k=0$ implies the near-horizon geometry is static and therefore henceforth we assume $k \neq 0$. By rescaling the coordinate $x$ we will set $k=1$.

Let us now return to the Maxwell equations (\ref{u1max}). The $x$ component allows one to solve for $B=E'$. This can then be substituted into the $\rho$ component to give
\be
\frac{\Gamma}{\sqrt{\gamma}} \left( \frac{\Gamma E'}{\sqrt{\gamma}} \right)' +E=0 \; .
\ee
In the analysis of vacuum near-horizon geometries~\cite{KL2}, it proved essential to introduce the coordinate $\sigma$ defined by $\sigma' =\sqrt{\gamma}$. Note that $\sigma$ is a globally defined function~\cite{KL2} and can be used as a coordinate everywhere except the points where $m$ vanishes. In terms of this coordinate the remaining part of the Maxwell equation simplifies to
\be
\Gamma \frac{d}{d\sigma}(\Gamma \dot{E})+E=0
\ee
where for functions of $\sigma$ we write $\dot{f}=df/d\sigma$. This equation can be integrated by noting that it is a total derivative if one multiplies it by $\dot{E}$. The result is
\be
\label{1stint}
(\Gamma \dot{E})^2+E^2=e^2
\ee
where $e$ is an integration constant and we assume $e>0$ (otherwise $E=0$ and hence $\cF=0$ which corresponds to the vacuum case).

The rest of the analysis closely follows that of the vacuum case
in~\cite{KL2}. We will work directly in $(\sigma,x)$ coordinates and
as in the vacuum case it is convenient to define the function
$Q(\sigma) \equiv \Gamma \sigma'^2 = \Gamma \gamma$ so the metric on
$\mathcal{H}$ is \be \gamma_{ab}dx^adx^b =
\frac{\Gamma(\sigma)}{Q(\sigma)} d\sigma^2 +
\frac{Q(\sigma)}{\Gamma(\sigma)} dx^2 \; . \ee The method involves
both local and global arguments and therefore at this stage it is
worth noting the following.  Since $\sigma$ is a globally defined
function, compactness of $\mathcal{H}$ implies that  $d\sigma$ must
vanish at distinct maxima and minima and thus since $(d\sigma)^2 =
Q/\Gamma$, we learn that $Q$ must vanish at these two distinct
points. Further since $(d\sigma)^2 =m \cdot m$ we see that $Q$ can only vanish at the fixed points of $m$ and thus $Q$ must only have two zeros on $\mathcal{H}$. Therefore we see that $\sigma_1 \leq \sigma \leq \sigma_2$
with $Q>0$ inside this interval and vanishing at the end points.

Equation (\ref{Feq}) gives \be A_0 + \frac{Q}{2\Gamma^2}-
\frac{1}{2} \nabla^2 \Gamma = \Gamma \left( -\frac{E^2}{\Gamma^2} -
\dot{E}^2 + \Lambda \right) \ee and integrating over $\mathcal{H}$
shows that $A_0<0$ and so we define $C>0$ such that $A_0=-C^2$.
Noting that $\nabla^2 f = \frac{d}{d\sigma} \left( \frac{Q
\dot{f}}{\Gamma} \right)$, this equation then reads \be \label{eq1}
\frac{d}{d\sigma} \left( \frac{Q \dot{\Gamma}}{\Gamma} \right)
=-2C^2+\frac{Q}{\Gamma^2}+2\left( \frac{E^2}{\Gamma}+
\dot{E}^2\Gamma -\Lambda\Gamma \right) \; . \ee Now, consider the
$xx$ component of (\ref{Riceq}). Using the fact that $R_{xx}=
-\frac{1}{2} \nabla^2 \gamma + \frac{1}{2} \dot{\gamma}^2$, it gives
\be \label{eq2} \ddot{Q}-\frac{d}{d\sigma}\left( \frac{Q
\dot{\Gamma}}{\Gamma} \right)+ 2\left(\frac{E^2}{\Gamma}+
\dot{E}^2\Gamma +\Lambda \Gamma \right) +\frac{Q}{\Gamma^2}=0 \; .
\ee Combining equations (\ref{eq1}) and (\ref{eq2})  in such a way
to eliminate the $\nabla^2 \Gamma$ term implies \be \label{Qode}
\ddot{Q}+2C^2+4\Lambda \Gamma=0 \ee which is the same as in the
vacuum case. Using (\ref{1stint}), equation (\ref{eq1}) can be
written as \be \label{Qexpr1} Q= \Gamma^2 \frac{d}{d\sigma} \left(
\frac{Q \dot{\Gamma}}{\Gamma} \right) +2\Gamma \left( C^2 \Gamma
-e^2+\Lambda \Gamma^2 \right) \ee and differentiating this
expression with respect to $\sigma$ implies \be \label{Qexpr2}
\dot{Q}= Q\Gamma \frac{d^3 \Gamma}{d\sigma^3} +
\ddot{\Gamma}(2\dot{Q}\Gamma-Q\dot{\Gamma} )+ 2\dot{\Gamma}( C^2
\Gamma+\Lambda \Gamma^2 -e^2) \ee where we have used (\ref{Qode}) to
eliminate the $\ddot{Q}$ generated by the differentiation.
Subtracting equation (\ref{Qexpr1}) (times $\dot{\Gamma}$) from
equation (\ref{Qexpr2}) (times $\Gamma$) leads to: \be \label{Gammaode}
Q \frac{d^3 \Gamma}{d\sigma^3} + \left( \dot{Q} -
\frac{\dot{\Gamma}Q}{\Gamma} \right)\left( 2\ddot{\Gamma}-
\frac{\dot{\Gamma}^2}{\Gamma}- \frac{1}{\Gamma} \right) =0 \; . \ee
This equation is identical to the one encountered in the vacuum case
in~\cite{KL2} and thus we may solve it using the same technique.

This involves noticing that if we define $\mathcal{P}=
2\ddot{\Gamma}- \frac{\dot{\Gamma}^2}{\Gamma}- \frac{1}{\Gamma}$
then (\ref{Gammaode}) can be written as \be \dot{\mathcal{P}}=
\left( \frac{\dot{\Gamma}}{\Gamma}- \frac{2\dot{Q}}{Q} \right)
\mathcal{P} \ee and thus can be integrated to give $Q^2\mathcal{P}=
k \Gamma$ where $k$ is the integration constant. In~\cite{KL2} it
was shown that $Q^2\mathcal{P}$ is a globally defined function which
vanishes where $Q$ does and thus evaluating at these points we learn
that the constant $k=0$ and hence $\mathcal{P}=0$. The differential
equation $\mathcal{P}=0$ is easily integrated to give \be
\dot{\Gamma}^2+1= \beta \Gamma \ee where $\beta$ is an integration
constant which must be positive. There are two solutions to this
equation. One is $\Gamma=\beta^{-1}$ which implies $Q$ is a constant
and is thus incompatible with $\mathcal{H}$ being compact. The other
solution is \be \Gamma= \beta^{-1}+
\frac{\beta(\sigma-\sigma_0)^2}{4} \ee where $\sigma_0$ is an
integration constant which we will set to zero using the freedom
that $\sigma$ has only been defined up to an additive constant. We
can now integrate (\ref{Qode}) for $Q$ to get \be\label{QKN} Q=
-\frac{\beta \Lambda}{12} \sigma^4 - (C^2+2\Lambda
\beta^{-1})\sigma^2+c_1\sigma+c_2 \ee where $c_i$ are two
integration constants. Substituting back into (\ref{Qexpr1}) gives
\be c_2= 4\beta^{-3}(C^2\beta +\Lambda -e^2\beta^2) \; . \ee Finally
we may integrate (\ref{1stint}) (which determines the Maxwell field)
resulting in \be \label{E} E=  \frac{ \sigma e\cos\alpha - \left(
\beta^{-1}- \frac{\beta \sigma^2}{4} \right) e\sin \alpha}{ \Gamma}
\ee where without loss of generality we have chosen a sign for $E$
and $\alpha $ is a constant. The rest of the near-horizon equations
are now satisfied identically.

We now complete the global analysis of the horizon metric. The procedure follows the case with vanishing Maxwell field~\cite{KL2} closely. In general the horizon metric we have derived has conical singularities at $\sigma=\sigma_1,\sigma_2$, the zeros of $Q$. Although the precise details of the argument depend on whether $\Lambda$ vanishes or not~\cite{KL2}, in both cases the condition for simultaneous removal of these conical singularities is equivalent to $Q(\sigma)$ being even, i.e. $c_1=0$,  so $\sigma_2=-\sigma_1>0$. In this case, the horizon metric is regular with $\partial / \partial x$ vanishing at the endpoints $\sigma =\pm \sigma_2$. This is of course consistent with $\mathcal{H}$ having $S^2$ topology. \\

\noindent {\it Summary of non-static near-horizon geometries} \\ \\
We have shown that there is a unique axisymmetric non-static near-horizon geometry with a compact horizon section of $S^2$ topology. It is given by:
\bea\label{NSsol}
ds^2&=& \Gamma[ -C^2 r^2 dv^2 +2dvdr] + \frac{\Gamma d\sigma^2}{Q} + \frac{Q}{\Gamma} (dx+rdv)^2 \\
\cF &=& d[ E (r dv + dx)] \eea where \be\label{NSdata} \Gamma =
\beta^{-1}+ \frac{\beta\sigma^2}{4}, \qquad Q = -\frac{\beta
\Lambda}{12} \sigma^4 - (C^2+2\Lambda
\beta^{-1})\sigma^2+4\beta^{-3}(C^2\beta +\Lambda -e^2\beta^2) \ee
and $E$ is given by (\ref{E}), where $C,\beta,e>0$ and $\alpha$ are
constants. This near-horizon solution is invariant under the scaling
\be\label{EMscaling} C^2 \to K C^2, \qquad \beta \to K^{-1} \beta,
\qquad e \to K e, \quad \sigma \to K \sigma, \qquad (v,x) \to
K^{-1}(v,x) \ee
where $K>0$ is a constant. This allows one to fix one, or a combination of, the parameters to any desired value. Therefore, it is a 3-parameter family. The horizon is at $r=0$ and the coordinates on spatial sections of this ($v= \const$) are $(\sigma,x)$. The polynomial $Q(\sigma)$ must have a roots at $\pm \sigma_2$ and the coordinate ranges are $-\sigma_2 \leq \sigma \leq \sigma_2$ (with $Q>0$ inside this interval) and $x$ is periodically identified in such a way to remove the conical singularities at $\sigma=\pm\sigma_2$. Therefore $\partial / \partial x$ has fixed points at $\sigma=\pm\sigma_2$. We will now show that this near-horizon geometry is in fact identical to the near-horizon limit of extremal Kerr-Newman for $\Lambda=0$ and extremal Kerr-Newman-AdS for $\Lambda<0$.\\

\noindent {\it Proof of equivalence to Kerr-Newman-AdS} \\ \\
First consider $\Lambda<0$, so we set $\Lambda = -3g^2$. In this case the function $Q$ (\ref{NSdata}) is a quartic with a positive $\sigma^4$ coefficient. As argued above, regularity requires that $Q$ be even with at least two distinct roots at $\sigma = \pm \sigma_2$ and be positive for $-\sigma_2 < \sigma < \sigma_2$. Hence $Q$ must have two additional real roots at $\sigma = \pm \sigma_3$ with $0< \sigma_2 <\sigma_3$. Observe that our solution depends on three parameters which may be taken to be: $(\sigma_2,\sigma_3,e,\alpha)$ subject to the scaling symmetry. Note that
\begin{eqnarray}\label{rootcond1}
 \sigma_2^2 + \sigma_3^2  &=& \frac{4C^2}{\beta g^2} - \frac{24}{\beta^2} \\
\sigma_2^2\sigma_3^2 &=& \frac{16}{\beta^4 g^2}\left(C^2\beta - 3g^2 - e^2\beta^2 \right) \; ,
\end{eqnarray} which are equivalent to
 \bea (\beta \sigma_2)^2(\beta \sigma_3)^2-4(\beta
\sigma_2)^2-4(\beta \sigma_3)^2=48 - \frac{16e^2\beta^2}{g^2}, \label{cond1}\\
\label{cond2} (\beta \sigma_2)^2(\beta \sigma_3)^2-2(\beta
\sigma_2)^2-2(\beta \sigma_3)^2= \frac{8C^2 \beta}{g^2} - \frac{16e^2\beta^2}{g^2} \; .\eea Introducing positive parameters $(a,r_+,q)$ (which are invariant under the scaling (\ref{EMscaling}))
\begin{equation}
 a \equiv \frac{\sigma_2}{g\sigma_3}, \qquad r_+ \equiv \frac{2}{g\beta \sigma_3}, \qquad q \equiv e\beta r_+^2
\end{equation} (note that $ag < 1$) and eliminating $\sigma_2, \sigma_3$ from~(\ref{cond1}), we find $g^2r_+^2 < 1$ and
\begin{equation}\label{KNasq}
 a^2 = \frac{r_+^2(1 + 3g^2r_+^2)-q^2}{1 - g^2r_+^2}.
\end{equation} One may eliminate $\sigma_2,\sigma_3$ from (\ref{cond2}), and using~(\ref{KNasq}) it follows
\begin{equation}\label{betaCsq}
 \beta C^2 = \frac{1 + 6g^2r_+^2 + a^2g^2}{r_+^2} \ .
\end{equation} Then using the scaling freedom~(\ref{EMscaling}) to set
\begin{equation}
C^2 =  \frac{1 + 6g^2r_+^2 + a^2g^2}{\Xi(r_+^2+a^2)},
\end{equation} where $\Xi \equiv 1 - a^2g^2$, we find from~(\ref{betaCsq}) $\beta = \Xi (r_+^2+a^2)/r_+^2$. Substituting into the definitions of $(r_+,a)$ gives
\begin{equation}
 \sigma_3 = \frac{2r_+}{g\Xi(r_+^2+a^2)}, \qquad \sigma_2 = \frac{2r_+ a}{\Xi(r_+^2 + a^2)}.
\end{equation} Next, perform the coordinate change
\begin{equation}
 \phi = \frac{2ar_+ x}{(r_+^2+a^2)^2}, \qquad \cos\theta =
 \frac{\sigma}{\sigma_2} \; ,
\end{equation} so $0 \leq \theta \leq \pi$ uniquely parameterises the interval, which implies
\be \label{KNdata} Q= \frac{4r_+^2a^2 \sin^2\theta
\Delta_{\theta}}{\Xi^3(r_+^2+a^2)^3}, \qquad \Gamma=
\frac{\rho_+^2}{\Xi(r_+^2+a^2)}, \qquad k^{\phi} = \frac{2ar_+ }{(r_+^2+a^2)^2},
\ee where $\Delta_{\theta}=1-a^2g^2\cos^2\theta$ and $\rho_+^2= r_+^2 +a^2 \cos^2\theta$. It is straightforward to verify \be \label{KNgam}\gamma_{ab}dx^adx^b = \frac{\Gamma d\sigma^2}{Q}+ \frac{Q}{\Gamma} dx^2=
\frac{\rho_+^2 d\theta^2}{\Delta_{\theta}}+\frac{\sin^2\theta
\Delta_{\theta} (r_+^2+a^2)^2}{\rho_+^2 \Xi^2} d\phi^2 \; \ee
and it is easy to see that absence of conical singularities implies $\phi \sim \phi +2\pi$. Next, define
\begin{equation}
 q_e \equiv -q\sin\alpha \qquad q_m \equiv -q\cos\alpha \; .
\end{equation} We find
\begin{equation}
\label{KNE}
 E = \frac{1}{\rho_+^2\Xi(r_+^2+a^2)}\left[q_e(r_+^2 - a^2\cos^2\theta) - 2q_m r_+ a \cos\theta \right]
\end{equation} and hence the space-time gauge potential is
\begin{equation}
 \cA = \frac{1}{\rho_+^2\Xi}\left(q_e\left[\frac{(r_+^2 - a^2\cos^2\theta)rdv}{r_+^2+a^2} + r_+ a \sin^2\theta d\phi\right] - q_m\left[\frac{2r_+ a \cos\theta r dv}{r_+^2+a^2} + (r_+^2 + a^2)\cos\theta d\phi\right]\right)
\end{equation}
where we have performed a gauge transformation to ensure a good $a \to 0$ limit.
We have thus rewritten our near-horizon solution~(\ref{NSsol}), (\ref{NSdata}), (\ref{E}) in terms of the coordinates $(\theta,\ \phi)$ and parameters $(a,\ r_+, \ q_e, \ q_m)$ satisfying the constraint~(\ref{KNasq}). It is easy to check that this is exactly the same as the near-horizon limit of the extremal rotating Kerr-Newman-AdS$_4$~\cite{CK}\footnote{In fact, the $a \to 0$ limit is simply the static near-horizon geometry of extremal Reissner-Nordstrom-AdS$_4$.}.

Finally consider $\Lambda=0$. In this case the solution we have
derived is simpler. We have $c_2= 4(C^2-e^2\beta)/\beta^2$ and since
$Q=-C^2\sigma^2+c_2$ we must have $c_2>0$ and $\sigma_2=
\sqrt{c_2}/C$. Changing variables to: \be \cos\theta =
\frac{\sigma}{\sigma_2}, \qquad \phi = \frac{2(C^2-e^2\beta)^{1/2}C
x}{2-e^2\beta^2C^{-2}} \ee so $0 \leq \theta \leq \pi$ uniquely
parameterises the interval, and introducing the positive parameters
(which are invariant under the scaling (\ref{EMscaling})) \be r_+^2
\equiv \frac{1}{\beta C^2}, \qquad q \equiv \frac{e}{C^2}, \qquad a
\equiv \sqrt{ \frac{1}{\beta C^2}\left(1- \frac{e^2\beta}{C^2}
\right) }  \; ,\ee so $r_+^2=a^2+q^2$, shows that the near-horizon
solution specified by $(\Gamma, k^{\phi}, \gamma_{ab}, E)$ is given
by equations (\ref{KNdata}), (\ref{KNgam}) and (\ref{KNE}) with
$g=0$. This confirms that our $\Lambda=0$ solution is identical to
the near-horizon limit of extremal Kerr-Newman.

\section{Supersymmetric near-horizon geometries}
For $\Lambda=-3g^2$, the theory we have been considering is the
bosonic sector of minimal gauged supergravity with gauge coupling
$g>0$, or minimal ungauged supergravity for $g=0$. Since any supersymmetric
black hole must be extremal\footnote{This can be seen as follows. Supersymmetry implies the existence of a globally defined non-spacelike Killing vector field $V$ (this is constructed as a bilinear of the Killing spinor). It follows that $V$ must be null and tangent on the event horizon $\mathcal{N}$ of a supersymmetric black hole. Hence $V$ is normal to $\mathcal{N}$, i.e. $\mathcal{N}$ is a Killing horizon of $V$. Since $V^2\leq 0$ it follows that $d(V^2)=0$ on $\mathcal{N}$ (i.e. the function $V^2$ is at an extremum on $\mathcal{N}$). Therefore the surface gravity vanishes, i.e. the black hole is extremal.}, our classification of near-horizon
geometries of extremal black holes includes a classification of
near-horizon geometries of supersymmetric black holes. Therefore, we
will now identify the subset of (axisymmetric) near-horizon
geometries in these theories, which are supersymmetric. The strategy
is to derive the integrability conditions arising from the existence
of a Killing spinor in this supergravity and figure out the
constraints this imposes on the near-horizon geometries we have
derived. We will take $g>0$, as for $g=0$ a classification has been
previously given in~\cite{CRT2}, where the result is simply the
near-horizon geometry of extremal Reissner-Nordstrom $AdS_2 \times
S^2$.

Recall we showed that for $g>0$, a static near-horizon geometry must
be the direct product of AdS$_2$ with $S^2,\ T^2$ or $H^2$. The
supersymmetry conditions for these solutions have been already
considered in~\cite{ortin} where it was shown that only the $H^2$
case with $\Delta^2+\phi^2=g^2$ is supersymmetric and preserves half the supersymmetries.
We will now analyse the non-static near-horizon geometry.

In fact, rather than calculating the integrability conditions
explicitly for our geometry, we will employ the following trick. We
will show that the non-static near-horizon geometry we derived is in fact
related by an analytic continuation to the
Reissner-Nordstrom-Taub-NUT-AdS$_4$ solution. The conditions for this solution to be supersymmetric are given in~\cite{ortin}, which we will exploit.

To see this, note the non-static near-horizon geometry we have derived has an $SO(2,1)$ isometry with 3d orbits which are
circle bundles over $AdS_2$. Consider analytic continuations of this
geometry, of the kind introduced in~\cite{KLR2}, in such a way to map $AdS_2 \to S^2$. The result will be an
$SO(3)$ invariant metric with 3d orbits, hence the 4d NUT charge.

More explicitly, first it is convenient to introduce global coordinates on the
AdS$_2$ in which case our near-horizon solution reads \bea
ds^2 &= &\Gamma \left( -(1+C^2Y^2)dT^2 +\frac{dY^2}{1+C^2Y^2} \right) + \frac{\Gamma}{Q}d\sigma^2 + \frac{Q}{\Gamma}(d\chi+YdT)^2 \\
\cF &=& d[ E(d\chi+YdT)] \eea where $(Y,T,\chi)$ are defined in
terms of $(r,v,x)$ as in~\cite{KLR2}. For the sake of generality we
will leave the constant $c_1$ appearing in the polynomial
$Q$~(\ref{QKN}) arbitrary. Now, define the parameters: \be N^2
\equiv -\frac{1}{C^2 \beta},  \qquad M \equiv  \frac{c_1}{4NC^4},
\qquad z \equiv \frac{ie}{C^{2}}, \qquad p \equiv z\sin\alpha,
\qquad q \equiv -z\cos\alpha \ee and coordinates \be r \equiv
\frac{\sigma}{2NC^2}, \qquad \cos\theta \equiv -iCY, \qquad \phi
\equiv CT, \qquad \tau \equiv -2iC^2 N \chi  \; . \ee It is then
easy to check that \bea
ds^2 = R^2 (d\theta^2 +\sin^2\theta d\phi^2) + \frac{R^2 dr^2}{\lambda} - \frac{\lambda}{R^2} (d\tau +2N\cos\theta d\phi)^2, \\
\cA_\tau = \frac{qr-Np}{R^2}+ \frac{p}{2N}, \qquad \cA_{\phi}=
\frac{\cos\theta}{R^2} \left( p(r^2-N^2)+2Nqr \right) \eea where
$R^2=r^2+N^2$ and $\lambda=g^2R^4 +(1+4g^2N^2)(r^2-N^2)-2Mr+z^2$.
This is the Reissner-Nordstrom-Taub-NUT-AdS solution exactly as given
in~\cite{ortin}.

As we showed earlier regularity of the horizon requires $c_1=0$ and
thus we need only consider the $M=0$ Reissner-Nordstrom-Taub-NUT-AdS solution above. In
this case the integrability conditions simplify a little and
are~\cite{ortin}:
\bea \label{int1} && g q N(1+4g^2N^2)=0 \\
\label{int2} &&(1 \pm 2gp + 4g^2N^2)(N^2(1+4g^2N^2) - p^2-q^2) = 0
\eea   where one may take either sign in (\ref{int2}). From these we
may deduce the integrability conditions for our non-static
near-horizon geometry by using the continuation above to convert to
our parameters. Recall the parameter ranges for our near-horizon
geometry are $\beta>0, C>0, e>0$. Note that
$1+4g^2N^2=1-\frac{4g^2}{C^2\beta}$ cannot vanish as if
$\beta=4g^2/C^2$ then our polynomial $Q(\sigma)$ has a single
stationary point at $\sigma=0$ and thus is not of the required form
for a compact horizon (see previous section). Since $N \neq 0$ we see that (\ref{int1})
implies $q=0$ and thus $\cos\alpha=0$ in terms of our near-horizon
geometry parameters. Now consider the second condition (\ref{int2}).
The factor $1 \pm 2gp + 4g^2N^2 = 1\pm 2gieC^{-2}\sin\alpha -
4g^2C^{-2}\beta^{-1}$ can only vanish if the real and imaginary parts
vanish separately. The imaginary part gives $\sin\alpha=0$ which is
not allowed since we also have $\cos\alpha=0$, and therefore we see
that this factor can never vanish. It follows that
$p^2=N^2(1+4g^2N^2)$. In terms of the near-horizon parameters this
is $e^2= \frac{C^2}{\beta}-\frac{4g^2}{\beta^2}$.

Therefore, to summarise the integrability conditions for the
existence of a Killing spinor for our non-static near-horizon
geometry are \be \label{susycond} \cos\alpha=0,  \qquad e^2=
\frac{C^2}{\beta}-\frac{4g^2}{\beta^2} \; .\ee Furthermore, we may
deduce the number of supersymmetries preserved by these near-horizon
geometries. From~\cite{ortin}, the $M=q=0$ and $p=N^2(1+4g^2N^2)$
RN-Taub-NUT-AdS solution actually preserves $1/2$ of the supersymmetries.
From the analytic continuation, it immediately follows that the
supersymmetric limit of our non-static near-horizon geometry
preserves the same number. In fact, as we show in the next section
this supersymmetric non-static near-horizon geometry is identical to
that of the known supersymmetric AdS$_4$ black hole~\cite{KP,CK}.

\section{Conserved charges}
We will now discuss what physical quantities of an extremal charged black hole can be computed from the near-horizon data alone, i.e. $(F,h_a,\gamma_{ab}, \Delta, \hat{\cF})$ which is data defined purely on $\mathcal{H}$.  In particular we consider solutions to (\ref{EOM}) and we will assume they are asymptotically flat for $\Lambda=0$ and asymptotically globally AdS for $\Lambda<0$, and in both cases $\cF \to 0$ (at a suitable rate) at infinity. We assume the black hole solution is axisymmetric, with $m$ being the associated rotational Killing field (which also leaves $\cF$ invariant) which we normalised to have period $2\pi$.

First consider the electric and magnetic charges:
\be
Q_e= \frac{1}{4\pi G} \int_{S^2_{\infty}} \star \cF, \qquad Q_m = \frac{1}{4\pi G} \int_{S^2_{\infty}} \cF \; .
\ee
By using Stokes' theorem $\int_{\Sigma} d\alpha= \int_{S^2_{\infty}} \alpha-\int_{\mathcal{H}} \alpha$ for a two form $\alpha$ and space-like hypersurface $\Sigma$, and noting that $(\star \cF)|_{\mathcal{H}}= \Delta \star_2 1$, we find
\be
Q_e= \frac{1}{4\pi G} \int_{\mathcal{H}} \Delta \sqrt{\gamma}, \qquad Q_m = \frac{1}{4\pi G} \int_{\mathcal{H}} \hat{\cF}
\ee
where we have used the Maxwell equation and Bianchi identity to show the integrals over $\Sigma$ vanish. It is thus clear that these physical quantities can be computed from the near-horizon data alone.

Now consider the angular momentum \be J= \frac{1}{16\pi G}
\int_{S^2_{\infty}} \star dm \; . \ee Using Stokes' theorem to
convert this to an integral over $\mathcal{H}$, we find a
non-trivial volume integral: \be J=\frac{1}{16\pi G}
\int_{\mathcal{H}} \star dm - \frac{1}{8\pi G}\int_{\Sigma} \star
R(m) \ee where we have used $\star d \star m =-2R(m)$ and
$R(m)_{\mu} \equiv R_{\mu\nu} m^{\nu}$. In~\cite{FKLR} it was shown
that \be J_H \equiv \frac{1}{16\pi G} \int_{\mathcal{H}} \star dm
=\frac{1}{16\pi G} \int_{\mathcal{H}} h \cdot m \sqrt{\gamma} \; .
\ee The new ingredient here is the volume integral over $\Sigma$,
which can be evaluated using Einstein's equations and the Maxwell
equations. The derivation is somewhat involved and very similar to
the proof of the Smarr relation (see e.g \cite{GMT} ) and thus we
omit details. However, we do point out that it is convenient to work
in a gauge where $\mathcal{L}_m\cA=0$ (of course we have
$\mathcal{L}_m \cF=0$ by assumption). In this gauge we
find\footnote{This expression is (as it should be) invariant under
residual gauge transformations, i.e. $\cA \to \cA+ d\lambda$ such
that $\mathcal{L}_m d\lambda=0$. To see this, note that the
constraint on $\lambda$ is equivalent to $m \cdot d\lambda $ being
constant, and further this constant must be zero since $m$ has fixed
points on $\mathcal{H}=S^2$.} \be J_{\Sigma} \equiv -\frac{1}{8\pi
G} \int_{\Sigma} \star R(m)= \frac{1}{4 \pi G}\int_{\mathcal{H}} (m
\cdot \cA) \star \cF \; . \ee Collecting these results we can
therefore write\footnote{It is worth noting that this expression is
also valid for Einstein-Maxwell-$\Lambda$ theory for $D>4$. Also
note that the same expression has been derived for rotating isolated
horizons~\cite{isolated}.} \be J= \frac{1}{16\pi G}
\int_{\mathcal{H}} \left( h \cdot m + 4(m\cdot \hat{A})\Delta
\right)\sqrt{\gamma} \ee which is valid in the gauge $\mathcal{L}_m
\hat{A}=0$.
We deduce that the angular momentum can also be computed
purely from the near-horizon data.

It is also natural to enquire whether the mass $M$ of an extremal
black hole can be computed from the near-horizon data alone. In
general this is not the case\footnote{In the special case of BPS
black holes one can determine the mass (see below). }, as one requires knowledge of how the stationary
Killing field is related to $V$ and $m$ which is not contained in
the near-horizon data. Similarly the angular velocity $\Omega_H$ and
the potential $\Phi_H$ cannot be deduced without asymptotic
information which one loses in the near-horizon limit. Of course one
can compute the area $A =\int_{\mathcal{H}} \sqrt{\gamma}$ from the
horizon metric $\gamma_{ab}$.

For the non-static near-horizon geometry (written in Kerr-Newman-AdS
coordinates and parameters) we have derived one can check that the
above integrals give \be Q_e = \frac{q_e}{G\Xi}, \qquad Q_m
=\frac{q_m}{G\Xi}, \qquad J=\frac{r_+(1+2g^2r_+^2+a^2g^2)a}{G\Xi^2},
\qquad A= \frac{4\pi (r_+^2+a^2)}{\Xi} \ee in agreement with the
corresponding quantities of the extremal Kerr-Newman-AdS black hole
(see e.g.~\cite{CK} and note in our conventions $a>0$ so $J>0$). For completeness we also give the mass of the
extremal Kerr-Newman-AdS, which cannot be computed from the
near-horizon geometry: \be M=\frac{r_+(1+2g^2r_+^2+a^2g^2)}{G\Xi^2}
\; . \ee

The theory we have been considering is the bosonic sector of gauged
minimal supergravity with gauge coupling $g$, whose solutions must
satisfy the BPS inequality $M \geq Z+gJ$ where $Z=
\sqrt{Q_e^2+Q_m^2}$~\cite{KP}. Such solutions admit Killing spinors
(i.e. are supersymmetric) only when the BPS condition is saturated
i.e. $M=Z+gJ$. For the extremal Kerr-Newman solution, it can be
checked this is satisfied if and only if $r_+^2=a/g$~\cite{KP}.
However, in~\cite{CK0} it was shown that this is not a sufficient
condition for the existence of a Killing spinor -- in addition, for
the Kerr-Newman solution one needs $q_m=0$. In terms of our derived
parameters this supersymmetric locus is $e^2 = C^2 \beta^{-1}-
4g^2\beta^{-2}$ and $\cos\alpha=0$. Observe that this agrees exactly
with the conditions we derived (\ref{susycond}) for the non-static
near-horizon geometry to be supersymmetric\footnote{This shows that
the most general supersymmetric non-static and axisymmetric
near-horizon geometry (with non-toroidal horizon topology) is in fact identical to that of the known
supersymmetric black hole.}. However, it is more interesting to
write this in terms of the physical quantities: \be
\label{constraint} J|Q_e|=g(G^2Q_e^4-J^2), \qquad Q_m=0 \; . \ee The
supersymmetric black hole is thus a one-parameter family of
solutions. We note that for this solution, the area as a function of
the conserved charges takes on a particularly simple
form\footnote{Note that as $g \to 0$, from the constraint
(\ref{constraint}) $J\to 0$ (one cannot have $Q_e\to 0$ as then from
the BPS relation $M \to 0$), giving the correct answer for extremal
Reissner Nordstrom.}: \be A= \frac{4\pi J}{g|Q_e|} = \frac{4\pi
(G^2Q_e^4-J^2)}{Q_e^2} \; .\ee This leads to a very simple expression
for the Hawking-Bekenstein entropy of the supersymmetric AdS$_4$
black hole. This is analogous to the expression found for
supersymmetric AdS$_5$ black holes~\cite{Kim}. Finally it is worth
noting that this supersymmetric black hole preserves $1/4$ of the
supersymmetries~\cite{CK0}. From the previous section we deduce that
its near-horizon geometry in fact preserves $1/2$ of the
supersymmetries. We thus find that supersymmetry is enhanced in the
near-horizon limit, as has been observed in five-dimensional gauged
supergravities~\cite{Sinha,KLR1,KL1}.

\section{Discussion}
In this note we have shown that the near-horizon geometry of any
rotating, (globally) AdS$_4$ extremal black hole in Einstein-Maxwell
theory must be given by that of the known extremal Kerr-Newman-AdS
black hole. We exploited the result of~\cite{HI}
which show that such a solution must be axisymmetric. Therefore we
assumed axisymmetry (and non-toroidal horizon topology), which is enough to allow a complete
classification of near-horizon geometries. This is a first step towards
proving a full uniqueness theorem for the extremal
Kerr-Newman-AdS$_4$ black hole. We have also shown (with no assumptions) that a static
near-horizon geometry must be a direct product of AdS$_2$ and
$S^2,T^2,H^2$. Since we are mainly interested in asymptotically
globally AdS$_4$ black holes, topological censorship then only
allows the $AdS_2 \times S^2$ solution. This latter solution is
simply the near-horizon geometry of extremal
Reissner-Nordstrom-AdS$_4$ which is the sub-case of the extremal
Kerr-Newman-AdS$_4$ black hole with $J= 0$.

One might be tempted to conclude that our combined results show that
the near-horizon geometry of any extremal black hole rotating or
otherwise must be given by that of the extremal Kerr-Newman-AdS
black hole (for some $J$, possibly vanishing). However this is not
quite correct, as it has not been shown that a non-rotating AdS
black hole must be either static or axisymmetric (the analogous
statement for asymptotically flat non-extremal black holes has been
shown~\cite{SW,W}). There is thus a potential gap, corresponding to
a non-axisymmetric, non-rotating, non-static extremal black hole
with a non-static near-horizon limit (which has no axisymmetry).
While it is tempting to dismiss such a possibility, we point out
that in an analysis of gravitational perturbations of higher
dimensional vacuum rotating AdS black holes~\cite{KLR0}, it is
suggested that the endpoint of the superradiant instability (which
occurs when the rotation is sufficiently fast as is the case for the
extremal limit) is precisely a black hole with such properties.

The results and techniques used in this note turned out to be a
straightforward generalisation of those for the analogous 4d vacuum
problem~\cite{KL2}, to include the effect of a Maxwell field. One
may contemplate further generalisations such as coupling to uncharged scalar
fields as occurs in (un)gauged supergravity coupled to extra
multiplets. It turns out that the resulting coupled ODEs do not seem
to admit a simple method of solution. Of perhaps more interest is to
generalise the 5d vacuum classification of~\cite{KL2} to include a
Maxwell field, as would be required for classifying
non-supersymmetric near-horizon geometries in 5d minimal (un)gauged
supergravity. We are currently examining this problem~\cite{KLa},
which does not appear to be straightforward generalisation of the
vacuum case.
\section*{Acknowledgements}
HKK and JL are supported by STFC. We thank Harvey Reall for useful comments.

\end{document}